\documentclass[]{article}
\usepackage{bagrow}
\usepackage{amsmath}
\usepackage{amssymb}
\usepackage{amsthm}
\usepackage{graphicx}
\usepackage[]{newtxtext}
\usepackage[bigdelims]{newtxmath}
\usepackage{booktabs}
\usepackage[usenames,dvipsnames,svgnames,table]{xcolor}
\usepackage{xcolor}



\newcommand{\wkd}{Wikidata}
\newcommand{\cn}{ConceptNet}
\newcommand{\ipr}{IPRnet}

\begin{document}

\author[1,2]{Daniel Berenberg}
\author[3,2,*]{James P.~Bagrow}
\affil[1]{Department of Computer Science, University of Vermont, Burlington, VT, United States}
\affil[2]{Vermont Complex Systems Center, University of Vermont, Burlington, VT, United States}
\affil[3]{Department of Mathematics \& Statistics, University of Vermont, Burlington, VT, United States}
\affil[*]{\corrauthinfo{james.bagrow@uvm.edu}{bagrow.com}}

\title{Inferring the size of the causal universe: features and fusion of causal attribution networks}

\date{December 14, 2018}

\maketitle

\begin{abstract}\raggedright
Cause-and-effect reasoning, the attribution of effects to causes, is one of the most powerful and unique skills humans possess.
Multiple surveys are mapping out causal attributions as networks, but it is unclear how well these efforts can be combined. 
Further, the total size of the collective causal attribution network held by humans is currently unknown, making it challenging to assess the progress of these surveys.
Here we study three causal attribution networks to determine how well they can be combined into a single network.
Combining these networks requires dealing with ambiguous nodes, as nodes represent written descriptions of causes and effects and different descriptions may exist for the same concept.
We introduce NetFUSES, a method for combining networks with ambiguous nodes.
Crucially, treating the different causal attributions networks as independent samples allows us to use their overlap to estimate the total size of the collective causal attribution network. 
We find that existing surveys capture 5.77\% $\pm$ 0.781\% of the 
$\approx$293\,000 causes and effects estimated to exist, and $0.198\% \pm 0.174\%$ of the 
$\approx$10\,200\,000 
attributed cause-effect relationships.
\end{abstract}



%




Causality and causal reasoning are central questions of statistics, computer science, philosophy, and the cognitive sciences~\cite{pearl2009causality,bunge1960causality,Hume1738-HUMATO-3,kant1999critique}.
Recently, our understanding of causality has been revolutionized by new insights and large volumes of data~\cite{pearl2009causal,Shiffrin201608845}.
%
%
Large-scale data collection and surveys of large numbers of individuals are now possible at an unprecedented scale.
One class of new data enabled by the internet is very large-scale knowledge graphs, annotated semantic networks codifying large numbers of factual statements, events, and interrelationships between concepts~\cite{berners2001semantic,suchanek2007yago,zesch2008extracting}. 
Knowledge graphs allow generalization to new relationships using graph algorithms, and these algorithms have been applied to causal predictions~\cite{radinsky2013mining}.

In this work we study causal relationships encoded into networks. 
Nodes in these networks represent causes and effects, and directed links indicate cause-effect relationships.
Generally, these relationships are gathered by large-scale surveying of individuals, often as part of larger efforts to build 
general-purpose semantic networks~\cite{liu2004conceptnet,vrandevcic2014wikidata}, although dedicated experiments have also been conducted~\cite{cscw_ipr}.
As these relationships are contributed by individuals or groups of individuals, we refer to these networks as causal attribution networks.
Attribution theory, the study of how individuals perceive causality and attribute causes to effects, has long explored the cognitive biases that affect causal attribution~\cite{kelley1967attribution,taylor1975point,kelley1980attribution}.

\begin{figure}
\includegraphics[width=1\textwidth]
{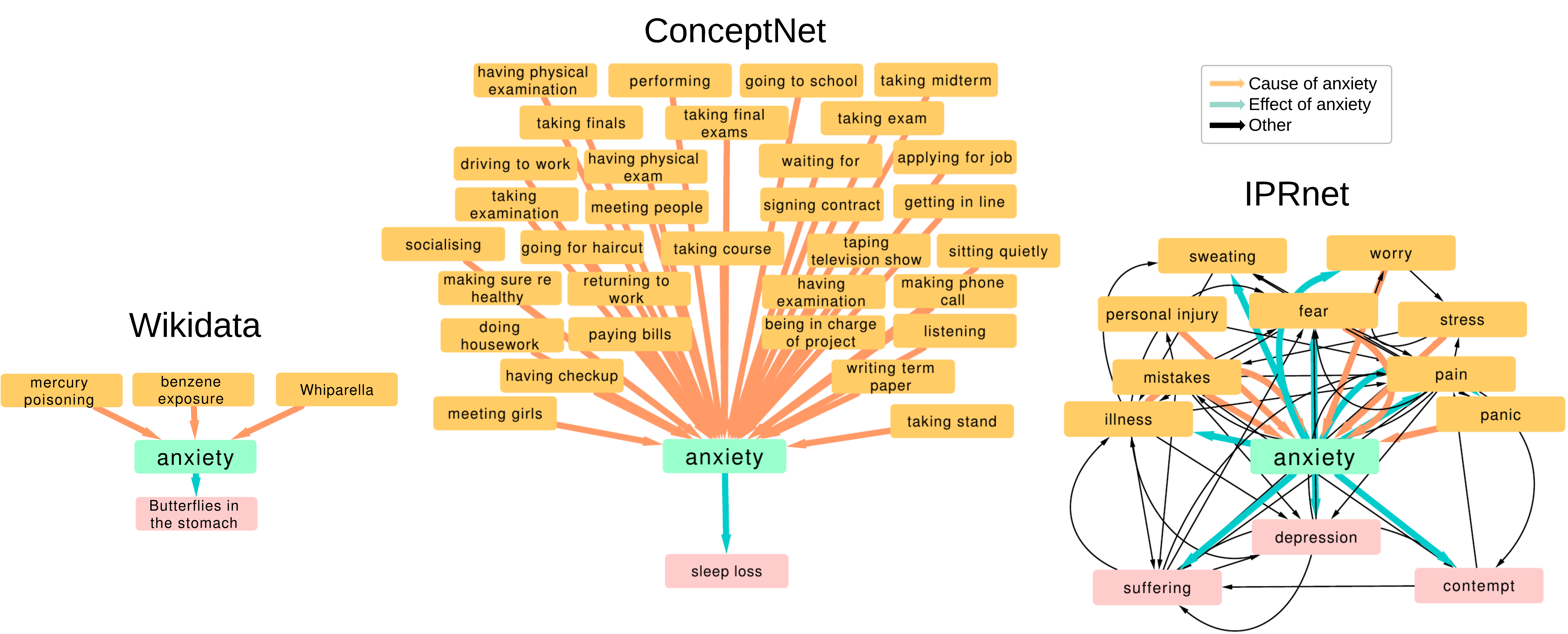}
\caption{
Causes and effects around `anxiety,' a term common to all three networks studied here.
This example illustrates similarities and differences of these networks, in particular the sparse, treelike structure of \wkd{} and \cn{} compared with the denser interlinking present in \ipr{}.
\label{fig:anxiety}
}
\end{figure}

We study causal attribution networks extracted from three sources, the collaboratively constructed knowledge graph ``Wikidata'', the long-running project ``ConceptNet,'' and ``IPRnet,'' a network built by members of a crowdsourcing platform to test a network data collection method called ``Iterative Pathway Refinement''~\cite{cscw_ipr}.
Figure~\ref{fig:anxiety} shows examples from all three networks, centered on ``anxiety'' a term common to all three.
Wikidata and ConceptNet encode other relationships, but we focus on causal relationships.
All three networks represent different efforts to explore the larger causal attribution network, and our goal is
to understand similarities and difference between these networks, and whether or not they can be fruitfully combined into a single, larger network.
A key challenge when combining these data is resolving ambiguity between entities or concepts: most causes and effects in these networks are originally identified only by short written descriptions, and it is possible to describe the same entity in many different ways.
Yet, overcoming this challenge to fuse multiple causal attribution networks together brings multiple benefits: it provides a common network dataset for researchers to study causality and attribution, and by measuring the overlap of different network samples we can estimate the size of the single underlying, incompletely observed causal attribution network.

The rest of this paper is organized as follows.
Section~\ref{sec:datamethods} describes the data collection and network and text analysis procedures, introduces a method called NetFUSES for fusing graphs with potentially ambiguous node identities, and discusses capture-recapture estimators for inferring the size of partially observed populations.
Section \ref{sec:results} compares the three causal networks using both network and text statistics (Sec.~\ref{subsec:descriptivestats}), then investigates how well these networks may be combined (Sec.~\ref{subsec:graphfusion}). Section \ref{subsec:inferringsizecausaluniverse} leverages these results to estimate the size of the underlying causal attribution network using capture-recapture methods.
Finally, we discuss our results in Sec.~\ref{sec:discussion}.

\section{Data and Methods}
\label{sec:datamethods}

\subsection{Causal attribution datasets}
\label{subsec:datasets}

In this work we compare causal attribution networks derived from three datasets. 
A causal attribution dataset is a collection of text pairs that reflect cause-effect relationships proposed by humans (for example, ``virus \emph{causes} sickness'').
These written statements identify the nodes of the network (see also our graph fusion algorithm for dealing with semantically equivalent statements) while cause-effect relationships form the directed edges (``virus'' $\to$ ``sickness'') of the causal attribution network.

We collected causal attribution networks from three sources of data: English \wkd{}~\cite{vrandevcic2014wikidata}, English \cn{} \cite{liu2004conceptnet}, and \ipr{}~\cite{cscw_ipr}. 
\wkd{} and \cn{}, are large knowledge graphs that contain semantic links denoting many types of interactions, one of which is causal attribution, while \ipr{} comes from an Amazon Mechanical Turk study in which crowd workers were prompted to provide causal relationships.
%
Wikidata relations were gathered by running four search queries on the Wikidata API (query.wikidata.org).
These queries searched for relations with the properties:
"has immediate cause", "has effect", "has cause", or "immediate cause of".
The first and third searches reverse the order of the cause and effect which we reversed back.
We discarded any Wikidata relations where the cause or effect were blank, as well as one ambiguous relation where the cause was "NaN".
ConceptNet attributions were gathered by searching the English ConceptNet version 5.6.0 assertions for ``/r/Causes/'' relations.
Lastly, IPRnet was developed in~\cite{cscw_ipr} which we use directly.

The three networks together contain
$23\,239$ causal links and
$19\,096$ unique terms,
of which there are 
$4\,265$ and $14\,831$ unique causes and effects, respectively.






\subsection{Text processing and analysis}
\label{subsec:NLPmethods}

\subsubsection*{Analyzing causal sentences}

Each node in our causal attribution networks consists of an English sentence, a short written description of an associated cause and/or effect. 
Text analysis of these sentences was performed using CoreNLP v3.9.2 and NLTK v3.2.2~\cite{coreNLP2014,nltk2009}.
We computed Part-of-Speech (POS) tags and identified (but did not remove) stop words for these sentences.
We used the standard Brown corpus as a text baseline for comparison.
Text processing procedures such as lemmatization or removal of casing were not performed in order to retain information for subsequent operations.
A small number of ConceptNet sentences contained 
`/n' and `/v' codes within the text denoting parts-of-speech tags; we removed these before applying our own POS tagger.
POS tagging of the causal sentences and the baseline dataset was performed using CoreNLP by tokenizing each input using the Penn Treebank tokenizer then applying the Stanford POS tagger.
This tagger uses Penn Treebank tags. 
We aggregated these 36 tags into NLTK's universal tagset which consists of a simpler set of 12 tags including NOUN, VERB, ADJ, and more.  
To simplify presentation, we chose to further collect all non-verb, non-noun, and non-adjective tags into an ``Other'' tag. 
Stop words were identified using NLTK's English stop words corpus. 

\subsubsection*{Vector representations for words and sentences}

Word vectors, or embeddings, are modern computational linguistics tools that project words into a learned vector space where context-based semantics of text are preserved, enabling computational understanding of text via mathematical operations on the corresponding vectors~\cite{bengio2003neural}.
Many different procedures exist for learning these vector spaces from text corpora~\cite{bengio2003neural,turian2010word,mikolov2013distributed,pennington2014glove}.
Document embeddings, or ``sentence vectors,'' extend word vectors, representing more complex multi-word expressions in a vector space of their own~\cite{conneau2017supervised}.
Given two nodes $i$ and $j$ with corresponding sentences $s_i$ and $s_j$ and sentence vector representations $\mathbf{v}_i$ and $\mathbf{v}_j$, respectively,
the vector cosine similarity 
$\frac{ \mathbf{v}_i \cdot \mathbf{v}_j }{ \| \mathbf{v}_i \| \| \mathbf{v}_j \| }$
is a useful metric for estimating the semantic association between the nodes.
High vector similarity implies that textual pairs are approximately semantically equivalent and sentence vectors can better compare nodes at a semantic level than more basic approaches such as measuring shared words or n-grams.

We computed sentence vectors 
using TensorFlow~\cite{abadi2016tensorflow} v1.8.0 using the Universal Sentence Encoder v2, a recently developed embedding model that maps English text into a 512-dimensional vector space and achieves competitive performance at a number of natural language tasks~\cite{universalSentenceEncoder2018}. 
This model was pretrained on a variety of text corpora~\cite{universalSentenceEncoder2018}. 
The Universal Sentence Encoder was tested on several baseline NLP tasks including sentiment classification and semantic textual similarity, for each of which it performs with the highest accuracy.
Given the higher performance of the Universal Sentence Encoder with respect to textual similarity tasks, we elected to utilize it instead of other sentence encoding models including the character level CNN architecture used in Google's billion word baseline \cite{jozefowicz2016exploring}, and weighted averaging of word vector representations \cite{arora2016simple}.

\subsection{Graph fusion}
\label{subsec:graphfusionmethods}

Graph fusion takes two graphs $G_1=(V_1, E_1)$ and $G_2=(V_2,E_2)$ and computes a \emph{fused} graph $G = (V,E)$ by identifying and combining semantically equivalent nodes (according to some measure of similarity) within and between $V_1$ and $V_2$.
Graph fusion is closely related to graph alignment and (inexact) graph matching~\cite{EMMERTSTREIB2016180}, although fusion assumes the need to identify node equivalents both within and between the networks being fused, unlike alignment and matching which generally focus on uncovering relations \emph{between} $V_1$ and $V_2$.
Graph fusion is particularly important when a canonical representation for nodes, such as an ID number, is lacking, and thus equivalent nodes may appear and need to be combined.
This is exactly the case in this work, where each node is a  written description of a concept, and the same concept can be equivalently described in many different ways.

Here we describe Network FUsion with SEmantic Similarity (NetFUSES).
This algorithm computes the fused graph $G$ given a node similarity function $f: V \times V \to \{0,1\}$.
This $f$ should encode the semantic closeness between nodes $u$ and $v$, with $f(u,v) = 1$ for semantically equivalent $u$ and $v$ and $f(u,v) = 0$ for semantically non-equivalent $u$ and $v$.
We assume $f(u,v) = f(v,u)$ and $f(u,u) = 1$. 

To fuse $G_1$ and $G_2$ into $G$, first compute $F = \{f(u,v) \mid u,v \in V_1 \cup V_2 \}$. One can interpret $F$ as (the edges of) a \emph{fusion indicator graph} defined over the combined node sets of $G_1$ and $G_2$.
Each connected component in $F$ then corresponds to a subset of $V_1 \cup V_2$ that should be combined into a single node in $V$.
(One can also take a stricter view and combine nodes corresponding to completely dense connected components of $F$ instead of any connected components, but this strictness can also be incorporated by making $f$ more strict.)
%
Let $F_i$ indicate the connected component of $F$ containing node $i$. 
Abusing notation, one can also consider $F_i$ as representing the node in $G$ that the unfused node $i$ maps onto.
Lastly, we define the edges $E$ of the fused graph based on the neighborhoods of nodes in $V$.
The neighborhood $N_G(F_i)$ of each node $F_i \in V$ in the fused graph is the union of the neighborhoods of the nodes connected to $i$ in $F$:
for any node $i \in V_1 \cup V_2$, let $F_{i} = \{j \in V_1 \cup V_2 \mid i,j \mbox{~connected in~} F\}$ and
$
K(i; H) = \bigcup_{u \in F_i \cap H} N_{H}(u).
$
Then the neighborhood $N_G(F_i) = \{F_j \mid j \in K(i; G_1) \cup K(i; G_2) \}$  
defines the edges incident on $F_i$ in the fused graph and $G$ may now be computed.
Notice by this procedure that if an edge already exists in $G_1$ and/or $G_2$ between two nodes $u$ and $v$ that share a connected component in $F$, then a self-loop is created in $G$ when $u$ and $v$ are combined.
For our purposes these self-loops are meaningful, but otherwise they can be discarded.

\textbf{Semantic similarity~~}
In this work, each node $i$ is represented only by a short written sentence $s_i$, and two sentences $s_i \neq s_j$ may in fact be different descriptions of the same underlying concept. 
Hence the need for NetFUSES\@.
To relate two sentences $s_i$ and $s_j$ semantically, we rely upon recent advances in natural language processing that can embed words and multiword expressions into a semantically-meaningful vector space (see Sec.~\ref{subsec:NLPmethods}).
Let $\mathbf{v}_i$ be the ``sentence vector'' corresponding to $s_i$. 
Then define $f(i,j) = 1$ if 
$\frac{ \mathbf{v}_i \cdot \mathbf{v}_j }{ \| \mathbf{v}_i \| \| \mathbf{v}_j \| } > t$ and zero otherwise, for some parameter $t$.
In other words, we consider nodes $i$ and $j$ to be semantically equivalent when the cosine similarity between their vectors exceeds a given threshold $t$.
Our procedure in the main text determined $t=0.95$ as an approach threshold.





\subsection{Capture-recapture}
\label{subsec:capturerecapture}

Capture-recapture (also known as mark-and-recapture and recapture sampling) methods are statistical techniques for estimating the size of an unobserved population by examining the intersection of two or more independent samples of that population~\cite{nayak1988estimating,pollock1990statistical}. 
For example, biologists wishing to understand how many individuals of a species exist in an environment may capture $n_1$ individuals, tag and release them, then later gather another sample by capturing $n_2$ individuals. The more individuals in the second sample that carry tags, the more likely it is that the overall population $N$ is small; conversely, if the overlap in the samples is small, then it is likely that $N$ is large.
Capture-recapture is commonly used by biologists and ecologists for exactly this purpose, but it has been applied to many other problems as well, including estimating the number of software faults in a large codebase~\cite{nayak1988estimating} and estimating the number of relevant academic articles covering a specific topic of interest~\cite{webster2013estimating}.

The simplest estimator for the unknown population size $N$ is the Lincoln-Petersen estimator.
Assuming the samples generated are unbiased, meaning that each member of the population is equally likely to be captured, then the proportion of captured individuals in the second sample who were tagged should be approximately equal to the overall capture probability for the first sample, $n_1 / N \approx n_{12} / n_2$. Solving for $N$ gives the intuitive Lincoln-Petersen estimator
$\hat{N} = {n_1 n_2}/{ n_{12}}$,
for $n_{12} > 0$.
While a good starting point, this estimator is known to be biased for small samples~\cite{pollock1990statistical}, and much work has been performed to determine improved estimators, such as the well-known Chapman estimator~\cite{chapman1951some}. 

In this work we use the recently developed Webster-Kemp estimator~\cite{webster2013estimating}:
\begin{equation}
\hat{N} = \frac{\left(n_1-n_{12}+1\right)\left(n_2-n_{12}+1\right)}{n_{12}} + n_1 + n_2 - n_{12},
\label{eqn:websterkempN}
\end{equation}
which assumes (i) that one tried to capture as many items as possible (as opposed to predetermining $n_1$ and $n_2$ and capturing until reaching those numbers) and (ii) the total number of items found $n_1 + n_2 - n_{12} \gg 1$.
Webster and Kemp also derive the variance of this estimator:
\begin{equation}
\sigma^{2}_{\hat{N}} = \frac{(n_1-n_{12}+1)(n_2-n_{12}+1)(n_1+1)(n_2+1)}{n_{12}^{2}(n_{12}-1)},
\label{eqn:websterkempVarN}
\end{equation}
with $n_{12} > 1$,
allowing us to assess our estimate uncertainty. 
Equations~\eqref{eqn:websterkempN} and \eqref{eqn:websterkempVarN} are approximations when assuming a flat prior on $N$ but are exact when assuming an almost-flat prior on $N$ that slightly favors larger populations $N$ over smaller~\cite{webster2013estimating}.

\section{Results}
\label{sec:results}

Here we use network and text analysis tools to compare causal attribution networks (Sec.~\ref{subsec:descriptivestats}).
Crucially, nodes in these networks are defined only by their written descriptions, and multiple written descriptions can represent the same conceptual entity.
Thus, to understand how causal attribution networks can be combined,
 we introduce and analyze a method for fusing networks (Sec.~\ref{subsec:graphfusion}) that builds off both the network structure and associated text information and explicitly incorporates conceptual equivalencies. 
Lastly, in Sec.~\ref{subsec:inferringsizecausaluniverse} we use the degree of overlap in these networks as a means to infer the total size of the one underlying causal attribution network being explored by these data collection efforts, allowing us to better understand the size of collective space of cause-effect relationships held by humans.

\subsection{Comparing causal networks}
\label{subsec:descriptivestats}

We perform a descriptive analysis of the three datasets, comparing and contrasting their features and properties.
We focus on two aspects, the network structure and the text information (the written descriptions associated with each node in the network).
Understanding these data at these levels can inform efforts to combine different causal attribution networks (Sec.~\ref{subsec:graphfusion}).


\subsubsection*{Network Characteristics}
\label{subsubsec:networkcahracteristics}

\begin{table}
\centering
\caption{Network statistics across each dataset.
Abbreviations: d-directed, u-undirected, w-weak, s-strong.
\label{tab:netstats}}
{\footnotesize
\begin{tabular}{lrrr}
\toprule{}
& \wkd{} & \cn{} & \ipr{} \\
\midrule
Nodes                     &      5\,316 &      12\,741 &      394 \\
Edges                     &      3\,818 &      16\,796 &     1\,329 \\
Causes                    &      2\,558 &       1\,497 &      302 \\
Effects                   &      3\,022 &      11\,687 &      361 \\
Self-loops                &         7 &        126 &        1 \\
Reciprocated edges        &         7 &          7 &      214 \\
Feedback loops            &         0 &          1 &      986 \\
Feedforward loops         &         0 &         87 &     3541 \\
Average degree            &     1.436 &      2.637 &    6.746 \\
Clustering (u)            &  0.004194 &    0.00568 &   0.3804 \\
Assortativity             &  -0.06101 &    -0.0569 &  -0.1627 \\
Longest shortest path (d) &         7 &          8 &       15 \\
Longest shortest path (u) &        25 &         12 &       11 \\
Connected components (w)  &      1\,780 &        415 &        1 \\
Connected components (s)  &      5\,309 &      12\,732 &      153 \\
Giant component size (w)  &      1\,069 &      11\,940 &      394 \\
Giant component size (s)  &         3 &          3 &      242 \\

\bottomrule
\end{tabular}
}
\end{table}

\begin{figure}
\centering
\includegraphics[width=0.8\textwidth]{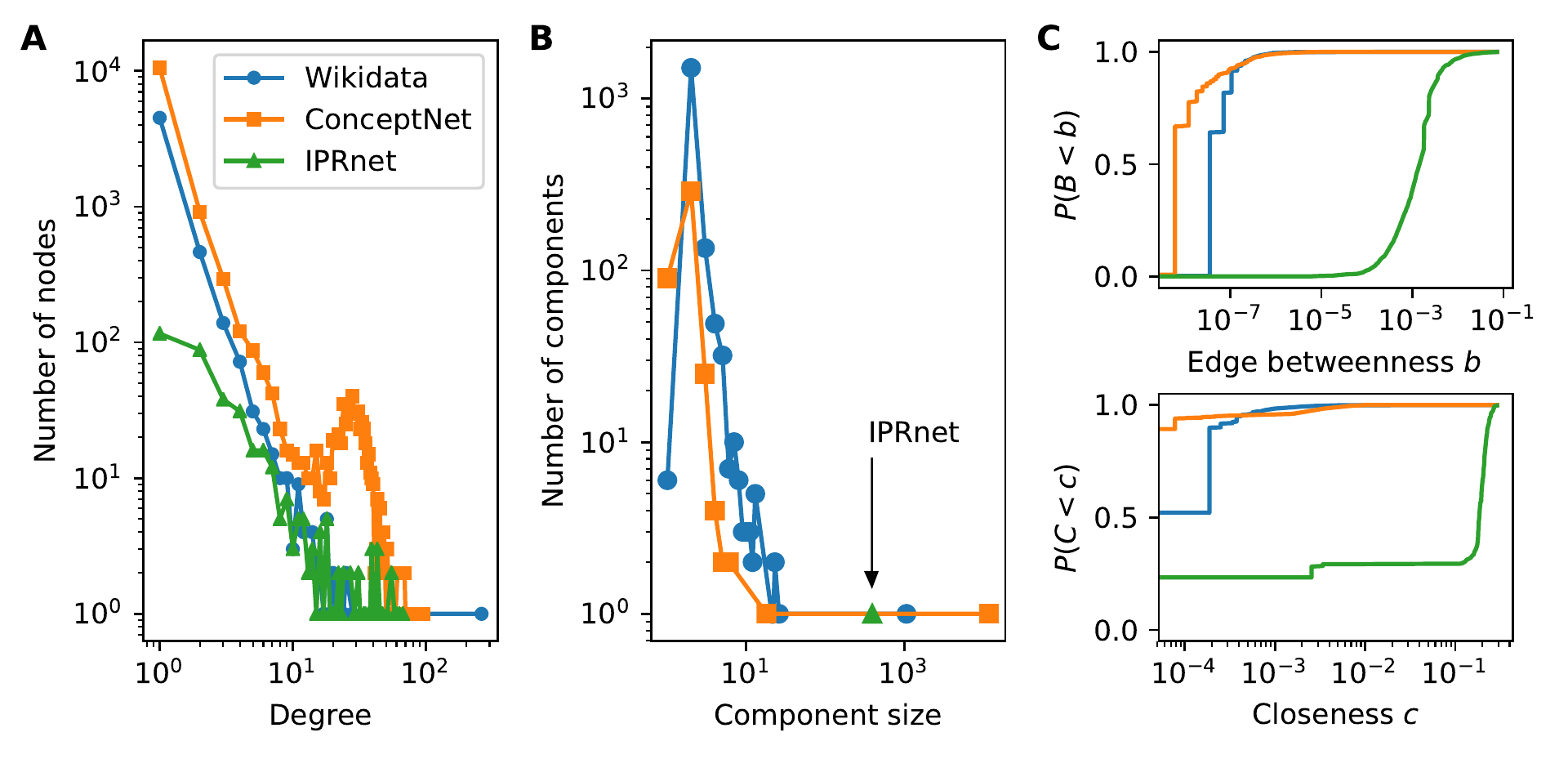}
\caption{Degree (\textbf{A}) and weakly connected component size (\textbf{B}) distributions for each data set as well as cumulative centrality distributions (\textbf{C}) for edge betweenness (top) and closeness (bottom) centrality measures. Note the interesting modality of high degree nodes in \cn{}.
\label{fig:connected_comps}}
\end{figure}

Table \ref{tab:netstats} and Fig.~\ref{fig:connected_comps} summarize network characteristics for the three causal attribution networks.
We focus on standard measures of network structure, measuring the sizes, densities, motif structure, and connectedness of the three networks.
Both \wkd{} and \cn{}, the two larger networks, are highly disconnected, amounting to collections of small components with low density. 
In contrast, \ipr{} is smaller but comparatively more dense and connected, with higher average degree, fewer disconnected components, and more clustering (Table~\ref{tab:netstats}).
All three networks are degree dissortative, meaning that high-degree nodes are more likely to connect to low-degree nodes. 
For connectedness and path lengths, we consider both directed and undirected versions of the network allowing us to measure strong and weak connectivity, respectively.
All three networks are well connected when ignoring link directionality, but few directed paths exist between disparate nodes in \wkd{} and \cn{}, as shown by the large number of strong connected components and small size of the strong giant components for those networks.

To examine motifs, we focus on feedback loops and feedforward loops, both of which play important roles in causal relationships~\cite{milo2002network,keil2006explanation}.
The sparse \wkd{} network has neither loops, while \cn{} has 87 feedforward loops and 1 feedback loop (Table \ref{tab:netstats}).
In contrast, \ipr{} has far more loops, 986 feedback and 3541 feedforward loops.

Complementing the statistics shown in Table \ref{tab:netstats}, Fig.~\ref{fig:connected_comps} shows the degree distributions (\ref{fig:connected_comps}A), distributions of component sizes (\ref{fig:connected_comps}B), and distributions of two centrality measures (\ref{fig:connected_comps}C).
All three networks display a skewed or heavy-tailed degree distribution.
We again see that \wkd{} and \cn{} appear similar to one another while \ipr{} is quite different, especially in terms of centrality.
One difference between \cn{} and \wkd{} visible in \ref{fig:connected_comps}A is a mode of nodes with degree $\sim 30$ within \cn{} that is not present in \wkd{}.

\subsubsection*{Text Characteristics}
\label{subsubsec:textcharacteristics}

Understanding the network structure of each dataset only accounts for part of the information.
Each node $i$ in these networks is associated with a sentence $s_i$, a written word or phrase that describes the cause or effect that $i$ represents.
Investigating the textual characteristics of these sentences can then reveal important similarities and differences between the networks.

\begin{figure}
\centering
\includegraphics[width=0.6\textwidth]
{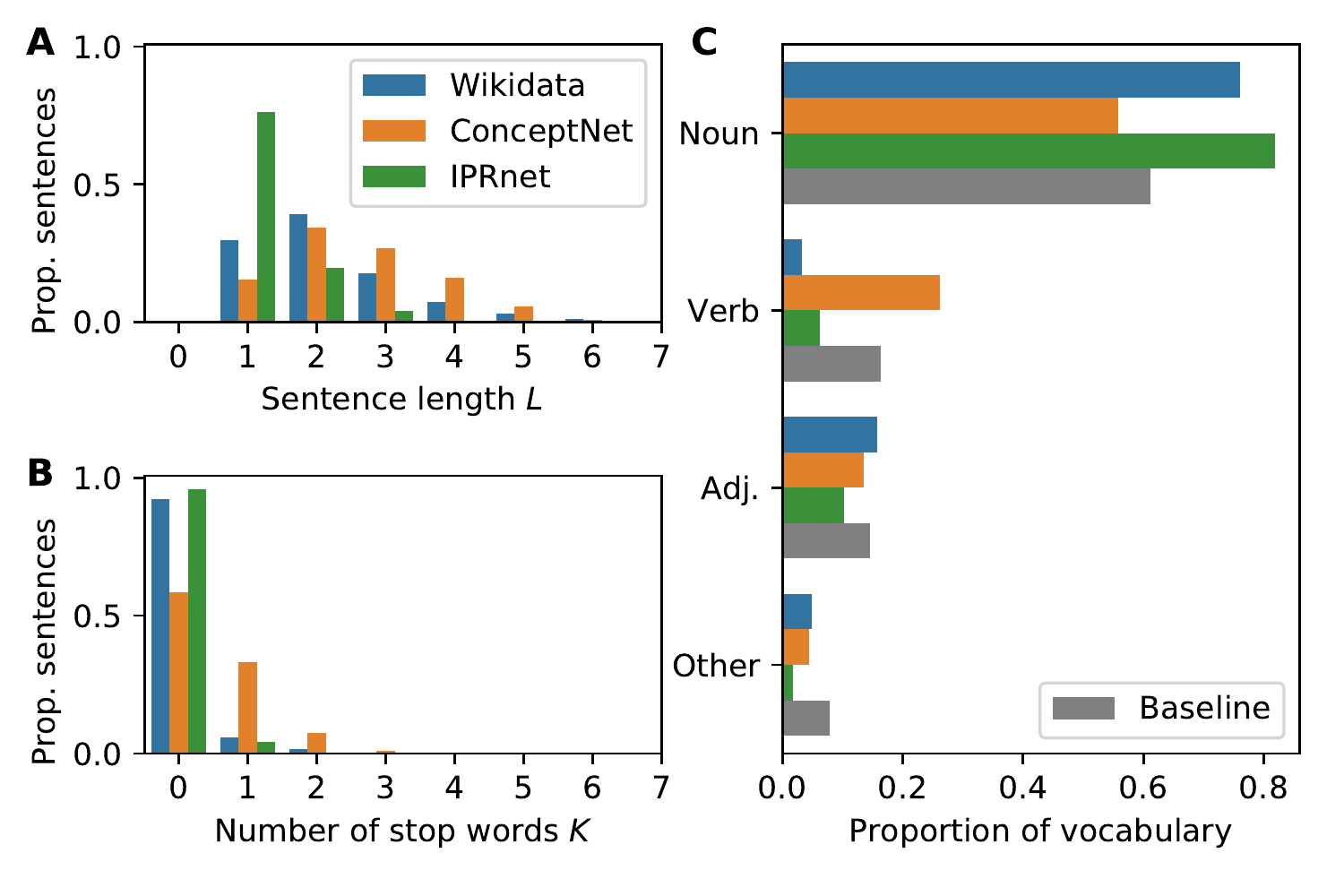}
\caption{%
Properties of causal attribution sentences across the  networks.
\lett{A} Distributions of total sentence lengths $L$ (number of words) for all unique sentences. 
\lett{B} Numbers of stop words per sentence.
\lett{C} Part of speech tags across words for each dataset, compared against a baseline POS tag distribution provided by the Brown corpus (grey).
Note that the horizontal axis in panel A has been truncated for clarity: $0.5\%, 0.02\%,$ and $0\%$ of \wkd{}, \cn{}, and \ipr{} sentences, respectively, have $L > 7$.
\label{fig:NLPdistrs}
}
\end{figure}

To study these sentences, we apply standard tools from natural language processing and computational linguistics (see Sec.~\ref{sec:datamethods}).
In Table \ref{tab:textstats} and Fig.~\ref{fig:NLPdistrs} we present summary statistics including the total text size, average length of sentences, and so forth, across the three networks.
We identify several interesting features. 
One, IPRnet, the smallest and densest network, has the shortest sentences on average, while ConceptNet has the longest sentences (Table \ref{tab:textstats} and Fig.~\ref{fig:NLPdistrs}A).
Two, \cn{} sentences often contain stop words (`the,' `that,' `which,', etc.; see Sec.~\ref{sec:datamethods}) which are less likely to carry semantic information (Fig.~\ref{fig:NLPdistrs}B).
Three, Wikidata contains a large number of capitalized sentences and sentences containing numerical digits. 
This is likely due to an abundance of proper nouns, names of chemicals, events, and so forth. 
These textual differences may make it challenging to combine these data into a single causal attribution network.

\begin{table}
\centering
\caption{Text statistics across each dataset.
\label{tab:textstats}}
{\footnotesize
\begin{tabular}{lrrr}
\toprule
& \wkd{} & \cn{} & \ipr{} \\
\midrule
Sentences   &     5\,316 &      12\,741 &      394 \\
Vocabulary size (unique words) &     5\,999 &       6\,822 &      399 \\
Text size (total words)        &    11\,960 &      33\,777 &      504 \\
Average sentence length             &     2.25 &      2.651 &    1.279 \\
Sentences with stop words            &      418 &       5\,296 &       16 \\
Sentences with numerical digits             &      533 &          8 &        0 \\
Sentences with punctuation          &      571 &        163 &        0 \\
Capitalized sentences               &     1\,849 &          0 &        0 \\

\bottomrule{}
\end{tabular}
}
\end{table}

We next applied a Part-of-Speech (POS) tagger to the sentences (Sec.~\ref{sec:datamethods}).
POS tags allow us to better understand and compare the grammatical features of causal sentences across the three networks, for example, if one network's text is more heavily focused on nouns while another network's text contains more verbs.
Additionally, POS tagging provides insight into the general language of causal attribution and its characteristics.
As a baseline for comparison, we also present in Fig.~\ref{fig:NLPdistrs}C the POS frequencies for a standard text corpus (Sec.~\ref{sec:datamethods}).
As causal sentences tend to be short, often incomplete statements, it is plausible for grammatical differences to exist compared with formally written statements as in the baseline corpus. 
For conciseness, we focus on nouns, verbs, and adjectives (Sec.~\ref{sec:datamethods}).
Nouns are the most common Part-of-Speech in these data,
especially for \wkd{} and \ipr{} that have a higher proportion of nouns than the baseline corpus (Fig.~\ref{fig:NLPdistrs}C).
\wkd{} and \ipr{} have correspondingly lower proportions of verbs than the baseline.
These proportions imply that causal attributions contain a higher frequency of objects committing actions than general speech.
However, \cn{} differs, with proportions of nouns and verbs closer to the baseline.
%
The baseline also contains more adjectives than \cn{} and \ipr{}.
Overall, shorter, noun-heavy sentences may either help or harm the ability to combine causal attribution networks, depending on their ambiguity relative to longer, typical written statements.


\subsection{Fusing causal networks}
\label{subsec:graphfusion}

These causal attributions networks are separate efforts to map out the underlying or latent causal attribution network held collectively by humans.
It is natural to then ask if these different efforts can be combined in an effective way.
Fusing these networks together can provide a single causal attribution network for researchers to study.
   
At the most basic level, one can fuse these networks together simply by taking their union, defining a single network containing all the unique nodes and edges of the original networks.
Unfortunately, nodes in these networks are identified by their sentences, and this graph union assumes that two nodes $i$ and $j$ are equivalent iff $s_i = s_j$. 
This is overly restrictive as these sentences serve as descriptions of associated concepts, and we ideally want to 
combine nodes that represent the same concept even when their written descriptions differ.
Indeed, even within a single network it can be necessary to identify and combine nodes in this way.
We identify this problem as \textit{graph fusion}. 
Graph fusion is a type of record linkage problem and is closely related to graph alignment and (inexact) graph matching~\cite{EMMERTSTREIB2016180}, but unlike those problems, graph fusion assumes the need to identify node equivalencies both within and between the networks being fused. 

We introduce a fusion algorithm, NetFUSES (Network FUsion with SEmantic Similarity) 
that allows us to combine networks using a measure of similarity between nodes (Sec.~\ref{sec:datamethods}).
Crucially, NetFUSES can handle networks where nodes may need to be combined even within a single network.
Here we compare nodes by focusing on the corresponding sentences $s_i$ and $s_j$ of the nodes $i$ and $j$, respectively, in two networks.
We use recent advances in computational linguistics to define a semantic similarity $S(s_i,s_j)$  between $s_i$ and $s_j$ and consider $i$ and $j$ as equivalent when $S(s_i,s_j) \geq t$ for some semantic threshold $t$.
See Sec.~\ref{sec:datamethods} for details.

To apply NetFUSES with our semantic similarity function (Sec.~\ref{sec:datamethods}) requires determining a single parameter, the similarity threshold $t$.
One can identify a value of $t$ using an independent analysis of text, but we argue for a simple indicator of its value given the networks: growth in the number of self-loops as $t$ is varied.
If two nodes $i$ and $j$ that are connected before fusion are combined into a single node $u$ by NetFUSES, then the edge $i\to j$ becomes the self-loop $u \to u$.
Yet the presence of the original edge $i \to j$ generally implies that those nodes are not equivalent, and so it is more plausible that combining them is a case of over-fusion than it would have been if $i$ and $j$ were not connected.
Of course, in networks such as the causal attribution networks we study, a self-loop is potentially meaningful, representing a positive feedback where a cause is its own effect.
But these self-loops are quite rare (Table~\ref{tab:netstats}) and we argue that creating additional self-loops via NetFUSES is more likely to be over-fusion than the identification of such feedback.
Thus we can study the growth in the number of self-loops as we vary the threshold $t$ to determine as an approximate value for $t$ the point at which new self-loops start to form.

\begin{figure}[t]
\centering
\includegraphics[width=0.5\textwidth]
{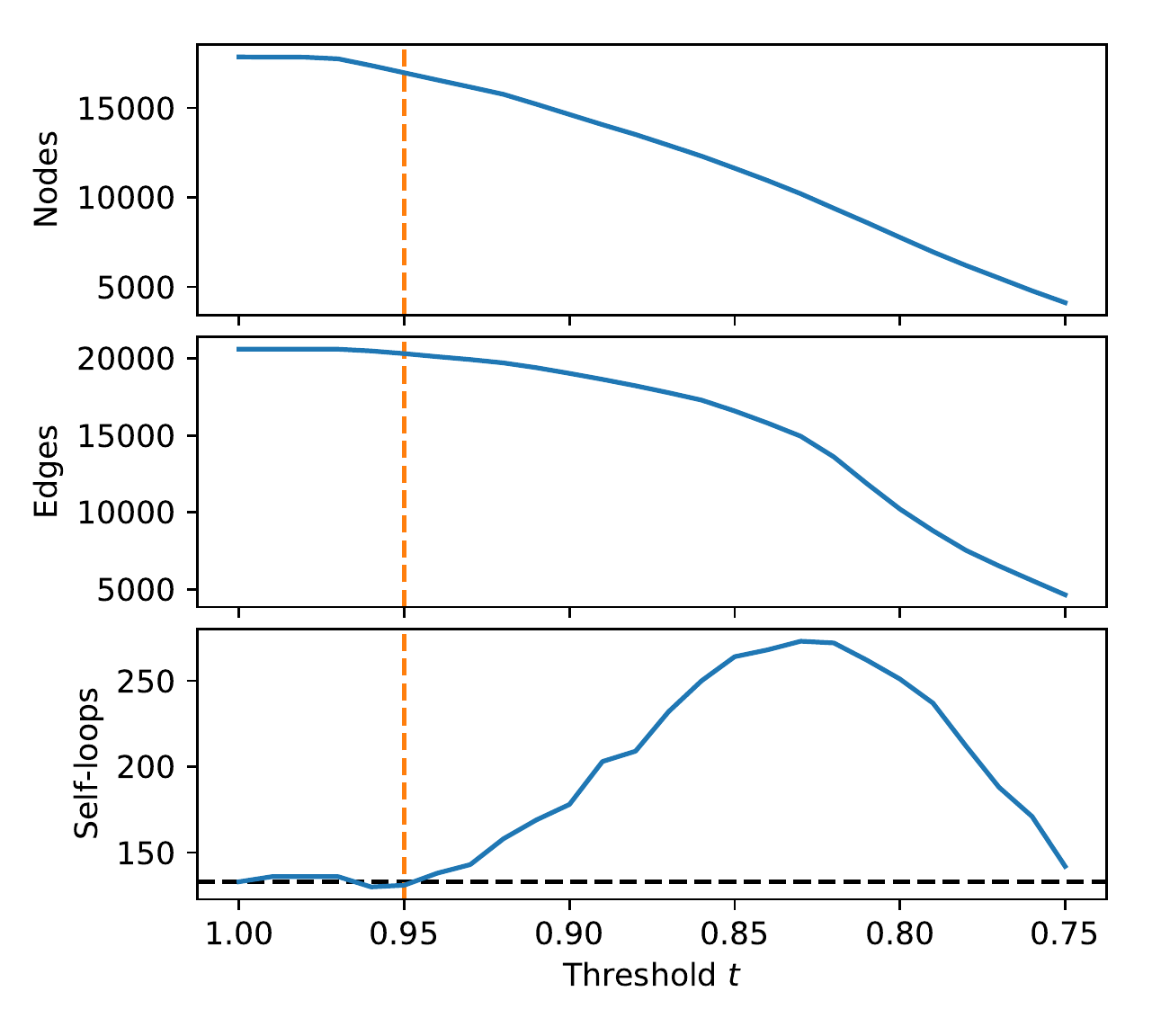}
\caption{
Statistics of fused \wkd{}--\cn{} networks across semantic similarity threshold values.
Monitoring the number of self-loops, we observe a relatively clear onset of over-fusion at a threshold of $t \approx 0.95$.
At this threshold, we observe a 4.95\% reduction in the number of nodes and a 1.43\% reduction in the number of edges compared with $t\geq1$.
\label{fig:fusion}
}
\end{figure}

Figure \ref{fig:fusion} identifies a clear value of the similarity threshold $t\approx 0.95$.
We track as a function of threshold the number of nodes, edges, and self-loops of the fusion of \wkd{} and \cn{}, the two largest and most similar networks we study.
The number of self-loops remains nearly unchanged until the level of $t = 0.95$, indicating that as the likely onset point of over-fusion. 
Further lowering the similarity threshold leads to growth in the number of self-loops, until eventually the number of self-loops begins to decrease as nodes that each have self-loops are themselves combined.
Thus, with a clear onset of self-loop creation, we identify $t = 0.95$ to fuse these two networks together.

\subsection{Inferring the size of the causal attribution network}
\label{subsec:inferringsizecausaluniverse}

These three networks represent separate attempts to map out and record the collective causal attribution network held by humans.
Of the three, IPRnet is most distinct from the other two, being smaller in size, denser, and generated by a unique experimental protocol. In contrast, \wkd{} and \cn{} networks are more similar in terms of how they were constructed and their overall sizes and densities. 

Treating \wkd{} and \cn{} as two independent ``draws'' from a single underlying network 
allows us to estimate the total size of this latent network based on their overlap.
(We exclude \ipr{} as this network is generated using a very different mechanism than the others.) 
High overlap between these samples implies a smaller total size than low overlap.
This estimation technique of comparing overlapping samples is commonly used in wildlife ecology and is known as capture-recapture or mark-and-recapture (see Sec.~\ref{subsec:capturerecapture}).
Here we use the Webster-Kemp estimator (Eqs.~\eqref{eqn:websterkempN} and \eqref{eqn:websterkempVarN}), but given the size of the samples this estimator will be in close agreement with the simpler Lincoln-Petersen estimator.

We first begin with the strictest measure of overlap, exact matching of sentences: node $i$ in one network overlaps with node $j$ in the other network only when $s_i = s_j$.
We then relax this strict assumption by applying NetFUSES as presented in Sec.~\ref{subsec:graphfusion}.

\wkd{} and \cn{} contain 12\,741 and 5\,316 nodes, respectively, and the overlap in these sets (when strictly equating sentences) is 208. 
Substituting these quantities into the Webster-Kemp estimator gives a total number of nodes of the underlying causal attribution network of $\hat{N} = 325\,715.4 \pm 43\,139.2$ ($\pm$ 95\% CI).
Comparing $\hat{N}$ to the size of the union of \wkd{} and \cn{} indicates that these two experiments have explored approximately 5.48\% $\pm$ 0.726\% of causes and effects.

However, this estimate is overly strict in that it assumes any difference in the written descriptions of two nodes means the nodes are different. Yet, written descriptions can easily represent the same conceptual entity in a variety of ways, leading to equivalent nodes that do not have equal written descriptions.
Therefore we repeated the above estimation procedure using \wkd{} and \cn{} networks after applying NetFUSES (Sec.~\ref{subsec:graphfusion}). 
NetFUSES incorporates natural language information directly into the semantic similarity, allowing us to incorporate, to some extent, natural language information into our node comparison.

Applying the fusion analysis of Sec.~\ref{subsec:graphfusion} and combining equivalent nodes within the fused \wkd{} and \cn{}, networks, then determining whether fused nodes contain nodes from both original networks to compute the overlap in the two networks, we obtain a new estimate of the underlying causal attribution network size of
$\hat{N} = 293\,819.0 \pm 39\,727.3$. 
This estimate is smaller than our previous, stricter estimate, as expected due to the fusion procedure, but within the previous estimate's margin of error.
Again, comparing this estimate to the size of the union of the fused \wkd{} and \cn{} networks implies that the experiments have explored approximately 5.77\% $\pm$ 0.781\% of the underlying or latent causal attribution network.

Finally, capture-recapture can also be used to measure the number of links in the underlying causal attribution network by determining if link $i\to j$ appears in two networks.
Performing the same analysis as above, after incorporating NetFUSES, provides an estimate of $\hat{M} = 10\,235\,150 \pm 8\,962\,595.9$ links. 
This estimate possesses a relatively large confidence interval due to low observed overlap in the sets of edges.
According to this estimate, $0.198\% \pm 0.174\%$ of links have been explored.


\section{Discussion}
\label{sec:discussion}

The construction of causal attribution networks generates important knowledge networks that may inform causal inference research and even help future AI systems to perform causal reasoning, but these networks are time-consuming and costly to generate, and to date no efforts have been made to combine different networks.
Our work not only studies the potential for fusing different networks together, but also infers the overall size of the total causal attribution network being explored.

We used capture-recapture estimators to infer the number of nodes and links in the underlying causal attribution network, given the \wkd{} and \cn{} networks and using NetFUSES and a semantic similarity function to help account for semantically equivalent nodes within and between \wkd{} and \cn{}.
The validity of these estimates depends on \wkd{} and \cn{} being independent samples of the underlying network. 
As with many practical applications of capture-recapture in wildlife ecology and other areas, here we must question how well this independence assumption holds. 
The best way to sharpen these estimates is to introduce a new causal attribution survey specifically designed to capture either nodes or links independently (it is unlikely that a single survey protocol can sample independently both nodes and links), and then perform this same survey multiple times.




NetFUSES is a simple approach to graph fusion, in this case building off advances made in semantic representations of natural language, although any similarity function can be used to identify semantically equivalent nodes as appropriate.
We anticipate that more accurate and more computationally efficient methods for graph fusion can be developed, but even the current method may be useful in a number of other problem domains.


\section*{Acknowledgments}

This material is based upon work supported by the National Science Foundation under Grant No.\ IIS-1447634.

{\singlespacing    

}

\end{document}